\documentclass[aps,onecolumn,showpacs,superscriptaddress,floatfix,preprint]{revtex4}
\usepackage{graphicx}
\usepackage{psfrag}
\usepackage{color}
\usepackage{float}
\usepackage{subfigure}
\usepackage{braket}
\usepackage{amsmath}
\usepackage{amsthm}
\usepackage{amsfonts}
\usepackage{amssymb}
\usepackage{mathrsfs}
\usepackage{epigraph}
\usepackage{fancyhdr}
\usepackage{wrapfig}
\usepackage{hyperref}
\usepackage{bm}

\begin{document}
\newcommand{\beq}{\begin{equation}}
\newcommand{\eeq}{\end{equation}}
\newcommand{\ben}{\begin{eqnarray}}
\newcommand{\een}{\end{eqnarray}}
\newcommand{\baq}{\begin{array}}
\newcommand{\eaq}{\end{array}}
\newcommand{\om}{(\omega )}
\newcommand{\bef}{\begin{}}
\newcommand{\eef}{\end{figure}}
\newcommand{\leg}[1]{\caption{\protect\rm{\protect\footnotesize{#1}}}}
\newcommand{\ew}[1]{\langle{#1}\rangle}
\newcommand{\be}[1]{\mid\!{#1}\!\mid}
\newcommand{\no}{\nonumber}
\newcommand{\etal}{{\em et~al }}
\newcommand{\geff}{g_{\mbox{\it{\scriptsize{eff}}}}}
\newcommand{\da}[1]{{#1}^\dagger}
\newcommand{\cf}{{\it cf.\/}\ }
\newcommand{\ie}{{\it i.e.\/}\ }
\newcommand{\he}{\mathcal{H}}
\setlength\abovedisplayskip{5pt}
\setlength\belowdisplayskip{5pt}

\title{A Quantum Biological Switch Based on Superradiance Transitions}
\author{D.~Ferrari}
\affiliation{Dipartimento di Matematica e
Fisica  Universit\`a Cattolica, via Musei 41, 25121 Brescia, Italy
}
\author{G.L.~Celardo}
\affiliation{Dipartimento di Matematica e
Fisica  Universit\`a Cattolica, via Musei 41, 25121 Brescia, Italy
}
\affiliation{ Interdisciplinary Laboratories for Advanced Materials Physics,
 via Musei 41, 25121 Brescia, Italy
\\
 Istituto Nazionale di Fisica Nucleare,  Sezione di Pavia,
via Bassi 6, I-27100,  Pavia, Italy}
\author{G.P.~Berman}
\affiliation{Theoretical Division, MS-B213, Los Alamos National Laboratories, Los Alamos, NM USA,
}
\author{R.T.~Sayre}
\affiliation{
Los Alamos National Laboratory and New Mexico Consortium, 202B Research Center,
Los Alamos, NM 87544, USA}
\author{F.~Borgonovi}
\affiliation{Dipartimento di Matematica e
Fisica  Universit\`a Cattolica, via Musei 41, 25121 Brescia, Italy
}
\affiliation{ Interdisciplinary Laboratories for Advanced Materials Physics,
 via Musei 41, 25121 Brescia, Italy
\\
 Istituto Nazionale di Fisica Nucleare,  Sezione di Pavia,
via Bassi 6, I-27100,  Pavia, Italy}

\begin{abstract}
A linear chain of connected electron sites with two asymmetric sinks, one attached to each end,
 is used as a simple model of quantum electron transfer
 in photosynthetic bio-complexes. For a symmetric initial population
 in the middle of the chain, it is expected that  electron transfer is mainly directed towards the strongest coupled sink.
However,  we show that quantum effects radically change this intuitive ``classical'' mechanism,
so that electron transfer can occur through the weaker coupled sink with   maximal efficiency.
Using this capability, we show how to design a quantum switch  that can transfer an electron
to the left or right branch of the chain, by changing the coupling to
the sinks.
The operational principles of this  quantum device can be understood in terms of  superradiance  transitions
  and  subradiant states.
This switching, being a pure quantum effect,
can be used as a witness  of  wave--like behaviour of excitations
in molecular chains.
When  realistic data are used for the photosystem II reaction center,
this quantum biological switch is shown to retain  its reliability, even at room temperature.

\end{abstract}

\date{\today}
\maketitle

\maketitle

\section{ Introduction.}
Understanding how biological systems transfer and store energy is a
basic energy science
challenge that can lead the design of new bio-nanotechnological devices
 \cite{Barber,Tachibana,Valov,SMGN,SC}.

Recent experiments on photosynthesis by several groups \cite{exp1,exp2,exp3,exp4,exp5,exp6,exp7,exp8,exp9},
have demonstrated the striking role of quantum coherence in the form of long lasting oscillations of the population of excitonic states in light harvesting complexes (LHC), at room temperature.
Even though there is no general consensus on the role quantum coherence plays in the
electron transfer (ET) efficiency ($\simeq 99 \%$) \cite{procon1,procon2,procon3,procon4,procon5,procon6}, there is no doubt that the models for exciton transport in the LHCs and primary charge separation in the reaction centers (RCs) should utilize quantum coherent effects.

The photosystem II (PSII) RC of many bacteria, plants and algae, where the primary charge separation occurs,
is arranged in two symmetric branches, even if only one of them is active for the ET. Different mechanisms which could be responsible for the asymmetry in the ET in the PSII RCs, and the related experiments, are discussed in   \cite{Scherer,Giorgi,Sayre1,Sayre2,Sayre3,pud1,pud2,arc1,
arc3,arc4} (see also references therein).

Here we do not address the question why only one
branch is active, but we use the PSII RC as a prototype for an artificial biological switch, able to drive the ET to the left or the right symmetric branch, by controlling the couplings to the sinks.

Primary charge separation in the RC can be modeled starting from a donor (a dimer, called the special pair where the excitation starts) and then including  the ET through different protein subunits,
(bacterio)chlorophylls and (bacterio)pheophytins, generally called chromophores.
This transfer occurs in a very short time (a few picoseconds).
On the other hand, the effective ET to the quinone occurs over
 a much longer time (a few microseconds) \cite{pud1,pud2,arc1,arc3,arc4,modrcq1,modrcq2,oscq}.
This allows one to consider a simplified model for the RC taking into account the short time dynamics
 corresponding to primary charge separation  (disregarding  long-time effects)
 by adding sinks through which the electron can escape the system. While there exists much experimental data for calculating both the energy levels and the couplings between the chromophores (typically dipole-dipole or Coulomb interactions), our aim here is to avoid non-essential technicalities which complicate the model, and concentrate on the main ideas of designing a quantum switching device based on the ET in the PSII RC.

The model can be described within the framework of an effective non-Hermitian Hamiltonian which takes into account the loss of electron probability to the sinks, one attached to each end of the system.
Even though this scheme has been  used in the past \cite{SR0,SR,srp,pud1,pud2},
to the best of our knowledge the intrinsic mechanism of superradiance and its relation 
to the ET has not been fully understood in bio-systems.
In bio-system superradiance has been considered as an effect  of the coupling with
the electromagnetic field  \cite{super} but actually it is a generic effect\cite{ZEL1} which can 
occur when a system is coupled to any continuum of states. Here we consider the phenomenon
of superradiance in transport induced by the coupling with the continuum of scattering
states as been discussed in \cite{SR0}. This point of view is related to the supertransfer
phenomenon discussed in Ref.~\cite{lst}.

\section{ The Model.}
\begin{figure}[t!]
\centering
\includegraphics[scale=1]{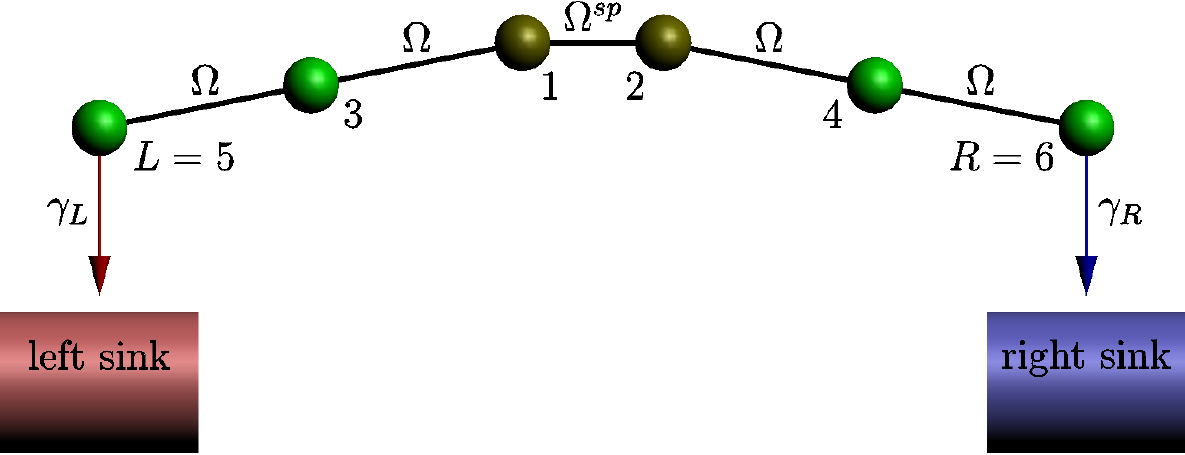}
\caption{Multimer model for the PSII RC.
The six sites have equal energies, $E_0 = 0$, and equal coupling constants, $\Omega$. The
pair $1-2$ is characterized by the coupling constant, $\Omega^{sp} > \Omega$.
Left and right ends of the chain are connected to the left and right sinks by
the coupling constants, $\gamma_{_L}$ and $\gamma_{_R}$, that
represent  couplings to
 the left and right continuum electron environment.
}
\label{l1}
\end{figure}

The model we consider consists of six sites divided into two symmetric branches, left and right, with two independent sinks attached at the ends. For simplicity, the energies of the sites are taken to be
equal, $E_0 = 0$, and the coupling between the nearest neighbor sites is constant ($\Omega$). The central pair of sites is allowed to have a  larger coupling constant, $\Omega^{sp} > \Omega$.
This very simple system was considered in the literature
(called the ``multimer" model \cite{multi1,multi2})
as a prototype model for the PSII RC, and
 is shown schematically in Fig. \ref{l1}. Despite its simplicity
(currently  more complicated models for the ET  have been introduced \cite{othtem1,othtem2}),
we believe it contains the essence of the process
we are modeling.
Later, in Sections III and IV, we will show that the results which follow from our model
maintain their validity in a large range of parameters when
more realistic models and thermal effects are considered.

The asymmetry in our model arises only from the different coupling strengths with the sinks,
 which can be understood as representing   continuum electron energy spectra.
These electron  environments are characterized by the transition rates to the left and to the right branches of the system.
We also choose  symmetric initial conditions for sites 1 and 2:
\begin{equation}
\rho(0)=\frac{1}{2}\left(\ket{1}+\ket{2}\right)\left(\bra{1}+\bra{2}\right).
\label{is}
\end{equation}
(Similar results can be obtained  for  symmetric mixed state,
$\rho(0)=\frac{1}{2}\left(\ket{1}\bra{1}+\ket{2}\bra{2}\right)$.)

An effective non-Hermitian Hamiltonian can be constructed as in \cite{srp} by
coupling the system to two different continuum electron reservoirs.
The probability flow into these continua is analyzed
with
an effective non-Hermitian Hamiltonian,  ${\cal H}$,
\begin{equation}
\mathcal{H}\equiv H_0 - i \gamma_{_L} W_L - i\gamma_{_R} W_R   =\left(
\begin{array}{cccccc}
0          &   \Omega^{sp}    &   \Omega             &0   &   0        &   0      \\
\Omega^{sp}     &   0         &  0        &   \Omega          &   0        &   0      \\
\Omega&   0         &   0             &   0          &   \Omega  &   0      \\
0 &   \Omega         &   0             &   0          &   0  &   \Omega      \\
0          &   0    &    \Omega  &   0          &   -i\gamma_{_L}/2        &   0      \\
0          &   0         &   0             &   \Omega         &   0   &   -i\gamma_{_R}/2
\end{array}
\right),
\end{equation}
\\ 
where $H_0$ is the Hamiltonian of the closed system, and $W_{L,R}$ take
 into account the coherent dissipation.
As one can see, in the site basis this corresponds  
to add  imaginary terms
to the end sites, $|L\rangle$
and $|R\rangle$, describing the loss of electron probability to the sinks.

The eigenvalues of $\mathcal{H}$ are complex numbers,
$E^{(k)} -i\Gamma^{(k)}/2$, where  $ \Gamma^{(k)}$ are the decay widths and
the evolution is described by the von Neumann equation,
\begin{equation}
\label{eq:masterq}
\frac{d\rho}{dt} = -\frac{i}{\hbar} \left( \mathcal{H}\rho -\rho\mathcal{H}^{\dagger} \right).
\end{equation}

\noindent
We also
 introduce the parameters, $\kappa_{_{L(R)}}$ and $q$,
\begin{equation}
\kappa_{_{L(R)}} \equiv \frac{\gamma_{_{L(R)}}}{2\Omega}, \qquad   q \equiv \frac{\kappa_{_L}}{\kappa_{_R}},
\label{kq}
\end{equation}
and the efficiencies of the ET to the sinks through the left ($|L\rangle$ site) and the right
($|R \rangle$ site) branches during  time, $T$:
\begin{equation}
\eta_{_{L(R)}}(T) = \frac{\gamma_{_{L(R)}}}{\hbar} \int_0^T dt\,
\bra{L(R)} \rho(t) \ket{L(R)}\\.
\label{eq:efficiency-def}
\end{equation}
 As shown in \cite{SR}, two superradiant transitions  with the corresponding
 formation of two superradiant (SR) states, are expected to occur, at
 
\begin{equation}
\left\{
\begin{array}{lll}
(ST_L) \qquad & \displaystyle \frac{\gamma_{_L}}{2\Omega} \simeq 1, \qquad & \Longrightarrow \qquad\kappa_{_L} \simeq 1 \\
&\\
(ST_R) \qquad & \displaystyle \frac{\gamma_{_R}}{2\Omega} \simeq 1, \qquad & \Longrightarrow \qquad \kappa_{_L} \simeq q
     \end{array}
\right.
\label{anak}
\end{equation}
\\
Strictly speaking, in \cite{SR} the ST has been found under the conditions
of a very large number of sites $N \gg 1$, and
$\Omega^{sp} = \Omega$.
We checked that the  STs occur
even for small $N$ values and 
in a large range of  $\Omega^{sp} \ne \Omega$ .

\begin{figure}[t!]
\centering
\includegraphics[scale=0.6]{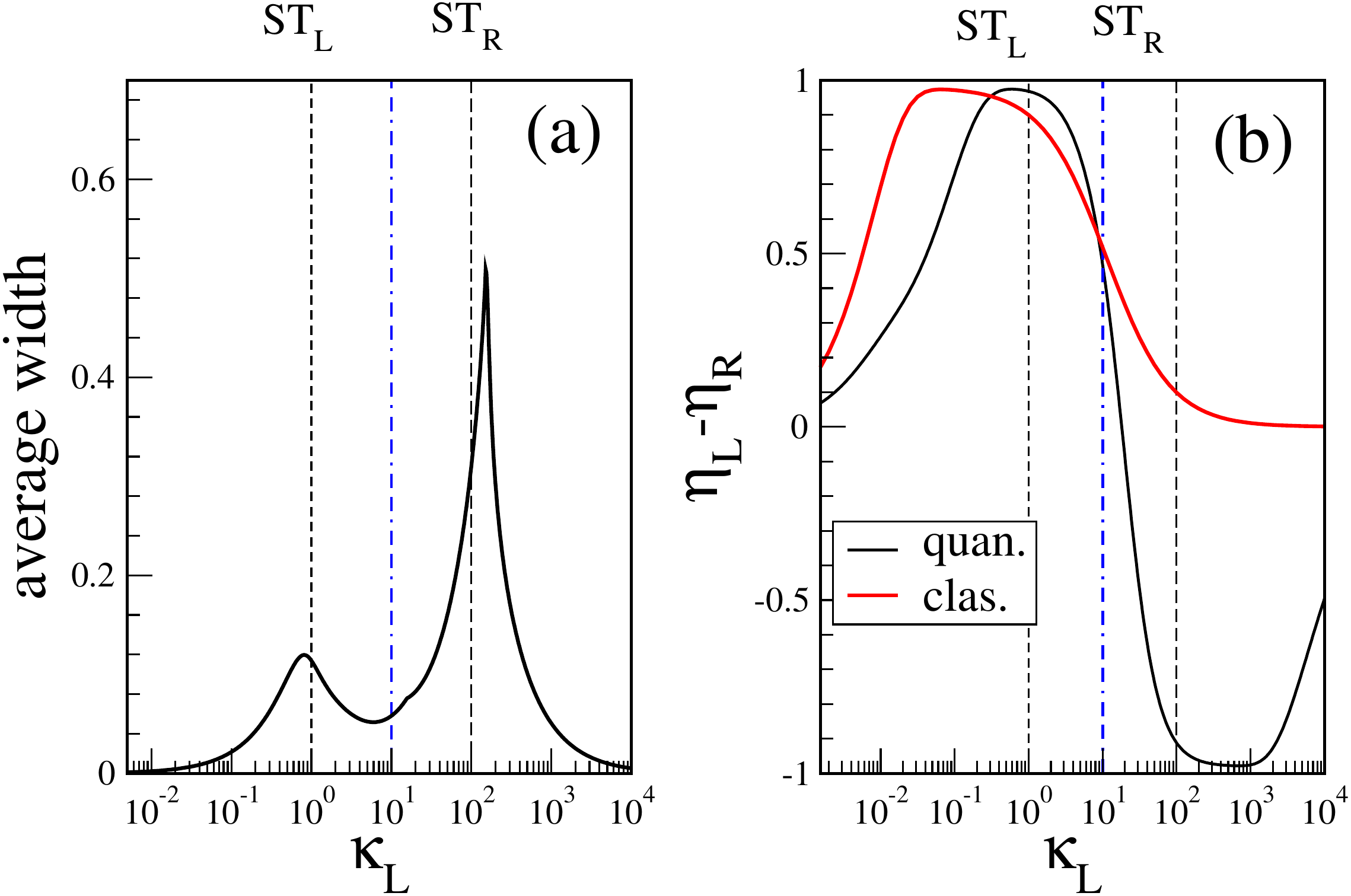}
\caption{(Color online) a) Average energy width of the $N-2$ eigenstates which
do not become superradiant
as a function of the effective coupling strength,
$\kappa_{_L}$, at fixed $q=\kappa_{_L}/\kappa_{_R} = 100$.
Average width  has also been renormalized by the average energy distance between levels, $D \simeq \Omega$.
b) Unbalanced left-right efficiency, $\eta_{_L} - \eta_{_R}$,
as a function of the effective coupling strength,
$\kappa_{_L}$, for the quantum case (black lower) and for the classical case (red upper).
Here $\Omega=100\  cm^{-1}$, ~$\Omega^{sp}=200 \ cm^{-1}$.
The efficiencies have been obtained by integrating over  $T=20 \ $ps.
In both panels, the vertical dashed lines represent, respectively, the left and the right STs ,
while the dashed-dotted central line indicates  the switching line, see text.
}
\label{l2}
\end{figure}

Let us analyze the physical picture  in which STs can be seen.
A small coupling with the continuum typically produces  level broadening, that is all
levels equally acquire
a width proportional to the strength of the opening. This `perturbative' argument
is valid up to a critical strength:  when  the widths of
neighboring levels overlap a `segregation' occurs. In other words
one energy level continues to have a width
proportional to  the opening (SR state) while all other levels (subradiant states)
are characterized by a  decay width proportional to the inverse of the opening strength.
This sharp transition has been called superradiant transition,
in analogy with the Dicke superradiance, since the SR state
 owns  a width $N$ times larger than the average width (if $N$ is the number of levels)
and it decays $N$ time faster than the other states. On the contrary,  the subradiants states,
in the limit of very large opening strength,  loose their widths and they do
not decay at all.  As shown in Ref. \cite{SR},
increasing $\kappa_L$ at fixed $q$ produces two STs, that can be observed by two peaks in the
average width of the $N-2$ subradiant states. (See Fig.~\ref{l2}a.)

We now  discuss the efficiency of the ET to the sinks under the condition that
the coupling to the left, $\gamma_{_L}$, is always larger than the
coupling to the right, $\gamma_{_R}$.
One might  expect
 that  the sink with the stronger coupling (stronger probability per unit of
time to escape the chain) will be the most efficient. But  what happens in this
  quantum  system is more complicated.
Indeed, Fig. \ref{l2}(b) shows that the unbalanced  efficiency,
$\eta_L-\eta_R$, as a function of the coupling strength, $\kappa_{_L}$, takes almost all values between -1 and 1, with two maxima close to the STs. In other words,  close to the left ST, $(ST_L, \kappa_{_L} \simeq 1 )$, the ET efficiency has a maximum
through
 the left branch, ($\eta_{_L} \simeq 1, \eta_{_R} \simeq 0 $), while close to the right  ST, $(ST_R, \kappa_{_L}  \simeq q )$, the ET efficiency has a maximum at the right branch, ($\eta_{_L}  \simeq 0, \eta_{_R} \simeq 1 $). This is an unexpected result
since the whole picture has been obtained under the condition $\kappa_L\gg\kappa_R$,
namely at the fixed ratio, $q=\kappa_{_L}/\kappa_{_R}=100 \gg 1$.

Therefore, the whole system can act as a probability  switch even if the left
coupling is much larger than the right coupling. How is this possible?
Before explaining these results and their intrinsic quantum nature, we
consider the ``classical'' behavior of this ET model.
The classical dynamics can  be modeled considering an incoherent hopping
among the sites, thus  described by
 a classical master equation with only diagonal density matrix elements (\emph{populations}):
\begin{equation}
\label{eq:class-mas-eq}
\frac{d\rho_{ii}}{dt} = \sum_k (T_{ik}\rho_{kk} - T_{ki}\rho_{ii})
-\left(
\frac{\gamma_{_L}}{\hbar} \delta_{Li} +
\frac{\gamma_{_R}}{\hbar} \delta_{Ri}
 \right)
\rho_{ii},
\end{equation}
where $\rho_{ii}$ is the probability at the $i$-th site, $T_{ik} = (H_0)_{ik}/\hbar $ is the transition rate
from the  $k$-th to the $i$-th sites,  and the last two terms represent the flow of probability through the left(L)
and  right(R) sinks.

The results of the classical dynamics, for the same model and symmetric initial conditions,
$\rho_{11}(0) = \rho_{22} (0) = 1/2$, are shown in Fig. \ref{l2}b (red curve).
They demonstrate
the absence of a  switch of transmission from the left to the right branch.
Indeed, one  always  finds
$\eta_{_L} > \eta_{_R}$, and  $\eta_{_L}  \simeq \eta_{_R} \simeq 1/2 $, for large values of $\kappa_{_L}$.
Up to some extent, this is in agreement with the intuitive interpretation: if the coupling
strengths to both sinks are extremely strong, particles will be absorbed by both sinks
with the same  efficiency. We can use  these results to define
  ``classical transport'' as one occurring through
the most strongly coupled branch, and  ``quantum transport'' as one occurring
through the weaker coupled branch.

These results  immediately raise  the following two
questions: How it can happen that  the ET occurs  through the weaker coupled branch?
Is it possible to estimate analytically  the ``switching point", $\kappa_L^{sw}$, located between two STs  at which $\eta_{_L} \simeq \eta_{_R}$?

To answer both questions, we  investigated the structure (localization
and decay width) of the eigenfunctions of the effective non-Hermitian Hamiltonian.
In the region between two STs where the switching occurs, there is
only one SR state. Even if its width is very large, it
becomes strongly localized around the  left sink, leaving  other subradiant states
approximately extended with no overlap with the left sink
(See Appendix A ).
This mechanism stops the ET through the left branch
and simultaneously  induces the ET through the right branch.

Let us  analytically estimate  the critical value, $\kappa_L^{sw}$, at
which the switching from left to right occurs.
Assuming that the switching occurs when the  partial decay
width to the left, $\Gamma_L$,  and to the right,  $\Gamma_R$, are equal, and
since
between the two STs, $\Gamma_L \propto 1/\kappa_L$
and $\Gamma_R \propto \kappa_L/q$ (see Appendix B and Ref.~\cite{albi})
one gets,
\begin{equation}
\frac {1}{\kappa_{_L}}
\simeq \frac{\kappa_{_L}}{q}
\qquad \Rightarrow  \qquad
 \kappa_{_L}^{sw} \simeq
 \sqrt{q}.
\label{ip}
\end{equation}
When the condition (\ref{ip}) is satisfied,
the
unbalanced ET efficiency is approximately zero.
This is verified directly in Fig. \ref{l2}b, where the theoretical dashed-dotted vertical line,
$\kappa_{_L}^{sw} \simeq \sqrt{q}$, is in a good agreement with the
switching point at which the unbalanced ET efficiency becomes zero.
 Note that in our case the same condition in
Eq.~(\ref{ip}) also
 defines the minimal decay width between the two STs, see  Fig. \ref{l2}a.
\begin{figure}[t!]
\centering
\subfigure[]
{\includegraphics[width=0.485\textwidth]{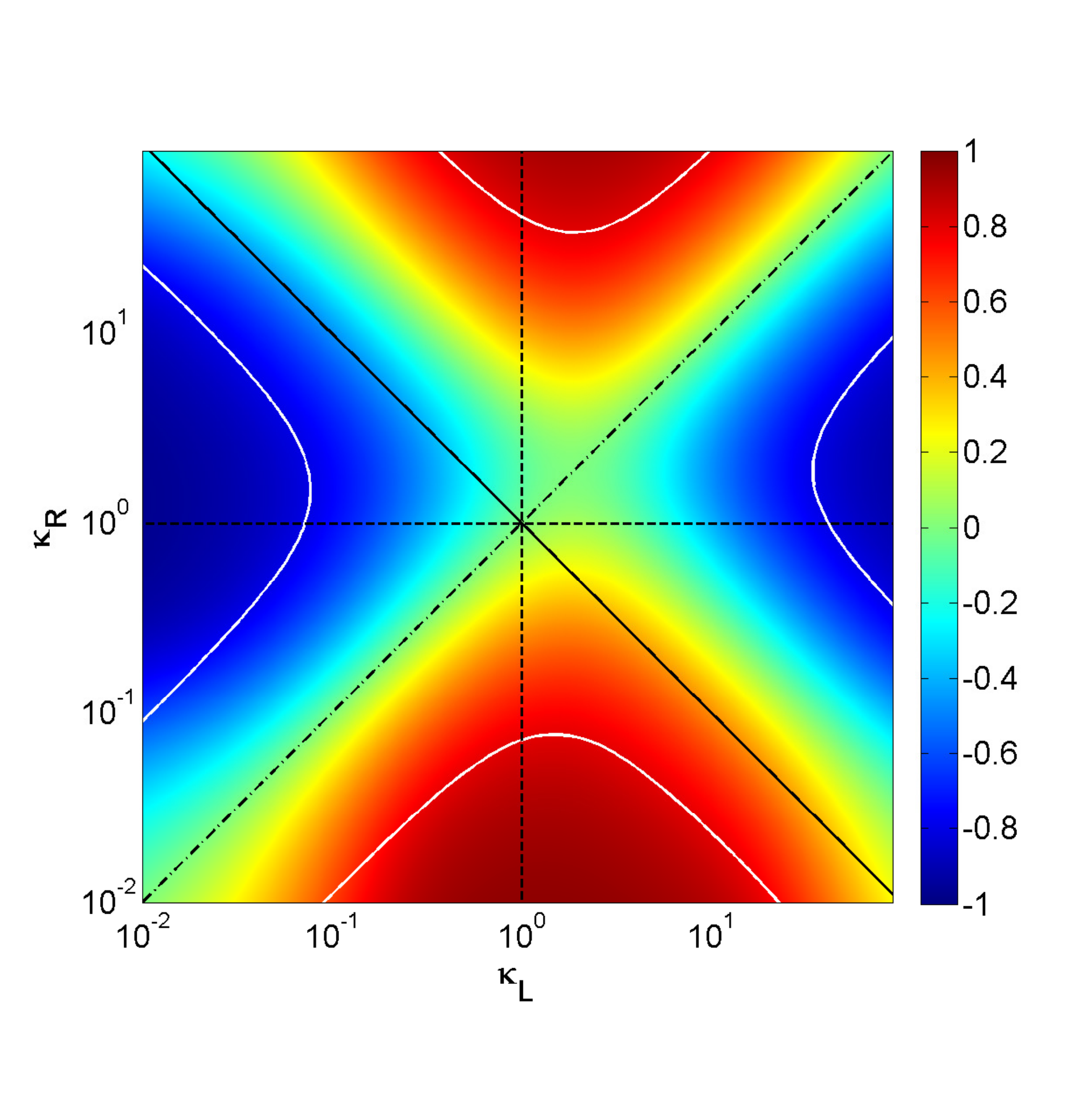}}
\subfigure[]
{\includegraphics[width=0.485\textwidth]{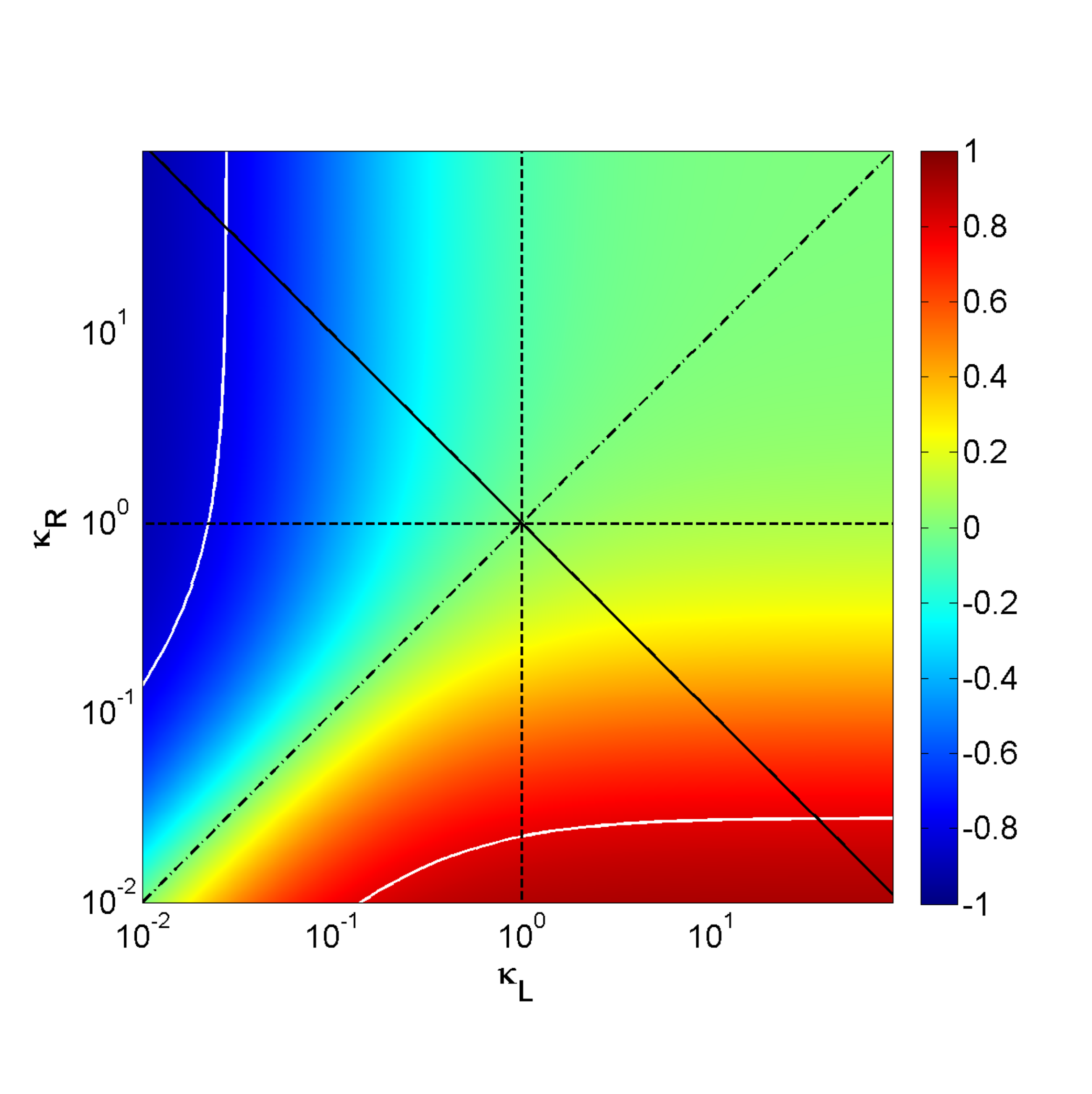}}
\caption{
The 2d contour plot, $\eta_L - \eta_R$, as a function of $\kappa_L=\gamma_{_L}/2\Omega$
and $\kappa_R=\gamma_{_R}/2\Omega$ in log-log scale
for the quantum  (a) and the classical case (b).
  The vertical and horizontal lines are,
respectively, the $ST_L$ and $ST_R$.
The two diagonals are the symmetry line $\kappa_L=\kappa_R$ (dashed)
and the switching line  $\kappa_L=1/\kappa_R$ (full).
White curves represent the $90$ \%  of the ratio between the two efficiencies ($\eta_L/\eta_R = 9$ or
$\eta_R/\eta_L = 9$).
}
\label{phases}
\end{figure}

One could ask whether these results are due to the strong asymmetry ($q=100$) used above.
The answer can be extracted from
the
 ``phase diagram'' of Fig.~\ref{phases},
in which  the unbalanced efficiency is shown
for all values of $\kappa_L$ and $\kappa_R$, and for both the quantum case (left panel) 
and the classical case (right panel), where the regions
of transport to the left have been indicated by red color and the transport to the right
branch by the blue color.
While the classical picture  shows that  to switch between  the red and the blue regions
it is necessary to cross the symmetry line, $\kappa_L = \kappa_R$, in the quantum world
a further possibility is given by crossing  the curve, $\kappa_L =  1/\kappa_R$.

\begin{figure}[t!]
\centering
\includegraphics[scale=0.6]{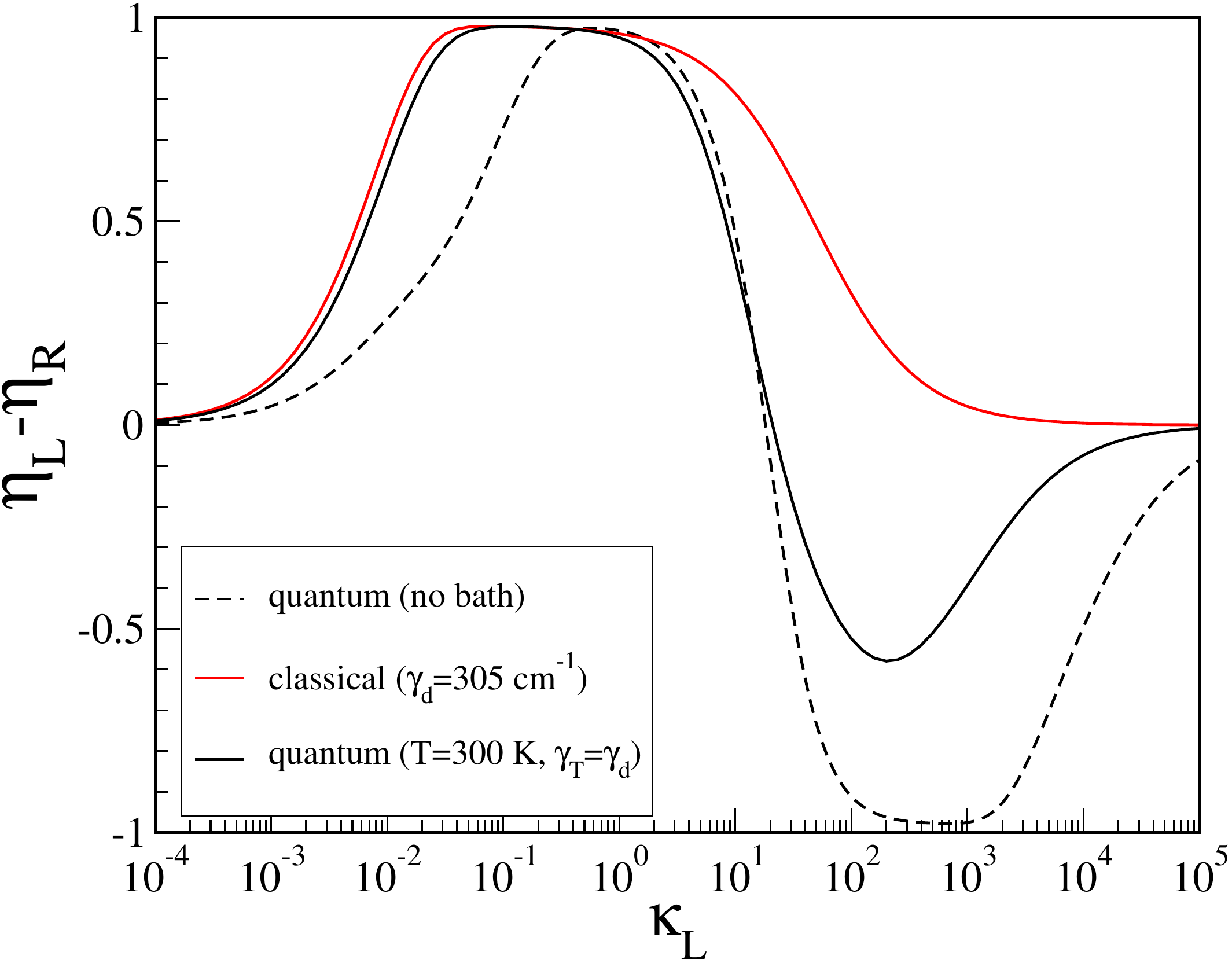}
\caption{(Color online)
Effects of a thermal bath. Unbalanced left-right ET efficiency,  $\eta_{_L} - \eta_{_R}$, as a
function of the effective coupling strength, $\kappa_{_L}$.
Black full line stands for quantum transport in presence of a thermal bath
at $T= 300 \ K$, $\omega_c = 150\ cm^{-1}$, $E_R = 35 \ cm^{-1}$, so to have
a homogeneous line broadening $\gamma_{_T} = 305.7 \ cm^{-1}$.
Dashed line is the quantum calculation did before (microcanonical).
Red line stands for the classical master equation  Eq.~\ref{eq:class-mas-eq}, with the
semiclassical rates, Eq.~\ref{rebe}, and $\gamma_{_d} = \gamma_{_T} = 305.7 \ cm^{-1}$.
Here $\Omega= \ 100 \ cm^{-1}$, $\Omega^{sp}= \
200 \ cm^{-1}$, and $q=\kappa_{_L}/\kappa_{_R}= \ 100$ fixed.
Left and right ET efficiencies have been obtained by integrating over   $20 \ $ps.
}
\label{ll3}
\end{figure}

\section{ Efficiency of the ET in the presence of a thermal bath.}
To demonstrate the
 robustness of the approach described above,  consider the interaction of our system with a
phonon bath at finite temperature. We use the thermal bath as in \cite{lloyd-bath}
whose dynamics is described by  the Lindblad master equation in the Born-Markov
and secular approximations.
More realistic models for the thermal bath can be found in the  literature \cite{othtem1,othtem2}.
This simpler approach  has already been used to describe the
FMO complex in a very similar framework \cite{lloyd-bath,srp}. Note
 that this approach leads to the relaxation of the populations, $\rho_{kk}$, to the Gibbs distribution.

We use the Lindblad-type master equation in the form,
\begin{equation}
\label{eq:master-lindblad}
\frac{d\rho}{dt} = -\frac{i}{\hbar} \left( \mathcal{H}\rho -\rho\mathcal{H}^{\dagger} \right)
+ L_p(\rho),
\end{equation}
where the action of the Lindblad operator, $L_p(\rho)$, on $\rho$ is described by Eq. (5) of
Ref. \cite{lloyd-bath}. (See Appendix C .)
In particular, we choose an exponential spectral  density dependent on two parameters,
the reorganization energy, $E_R$, and the cutoff frequency, $\omega_c$,  to be considered
together with  the temperature, $T$,  of the bath.

The thermal bath (interaction with phonons) produces an homogeneous line broadening \cite{classical},
proportional to both temperature and reorganization energy, and inversely proportional
to the cut-off frequency\cite{broad}
\begin{equation}
\label{rebe}
\gamma_{_T} = 2\pi \left(\frac{k T}{\hbar}\right) \left(\frac{E_R}{\hbar\omega_c}
\right).
\end{equation}
 In Fig.~\ref{ll3}, we plot the unbalanced left-right ET efficiency as a function of the coupling,
 $\kappa_{_L}$, at $T=300 $K, and reorganization energy and cut-off frequency chosen
in order to have an homogeneous line broadening $\gamma_{_T} = 305.7 \ cm^{-1}$ (full black curve).

As one can see, the first effect is that the  quantum switching due to
the superradiance is weakened but not suppressed by the thermal bath (compare
with the dashed curve which represents the same quantity in absence of the thermal bath
for the same value of $q= \kappa_{_L}/\kappa_{_R}=100$).

To have a close comparison with the classical model we consider the same classical master equation
as before, Eq.~(\ref{eq:class-mas-eq}), but with the transition rates, computed semiclassically
as in \cite{classical, srp},
\begin{equation}
\label{leg}
T_{ik} = \frac{ 2 \Omega_{ik}^2}{\hbar \gamma_{_d}}\left( 1+ \frac{\Delta E_{ik}^2}{\gamma_{_d}^2}
 \right)^{-1},
\end{equation}
where $\Omega_{ik}$  represents the energy coupling between the $i$-th and the  $k$-th sites, the
$\Delta E_{ik}$ are the energy differences between the two sites, and $\gamma_{_d}$ is
the  dephasing energy \cite{classical}.

To have a close comparison we put the dephasing energy $\gamma_{_d} = \gamma_{_T}$.
Results obtained from the classical master equation with the semiclassical rates Eq.~(\ref{leg})
are shown in  Fig.~\ref{ll3} as a red curve.

As one can see, the switching is absent in the classical 
model since 
the incoherent hopping transport gives at most $\eta_R \approx \eta_L = 1/2$.
The most interesting result is that the presence of thermal bath
has the opposite effects on  classical
and quantum transport:
the thermal bath weakens only the quantum transport (transport to the right weakest coupled
branch) while it leaves mainly unaffected the classical transport (left strongest coupled branch).
The different sensibility of  ET  to the dephasing
induced by the thermal bath 
is consistent with the quantum coherent nature of the switch.
This  opens the interesting possibility to use the switch effect as a  witness of 
quantum coherence  in molecular chains.

\begin{figure}[t!]
\centering
\includegraphics[scale=0.6]{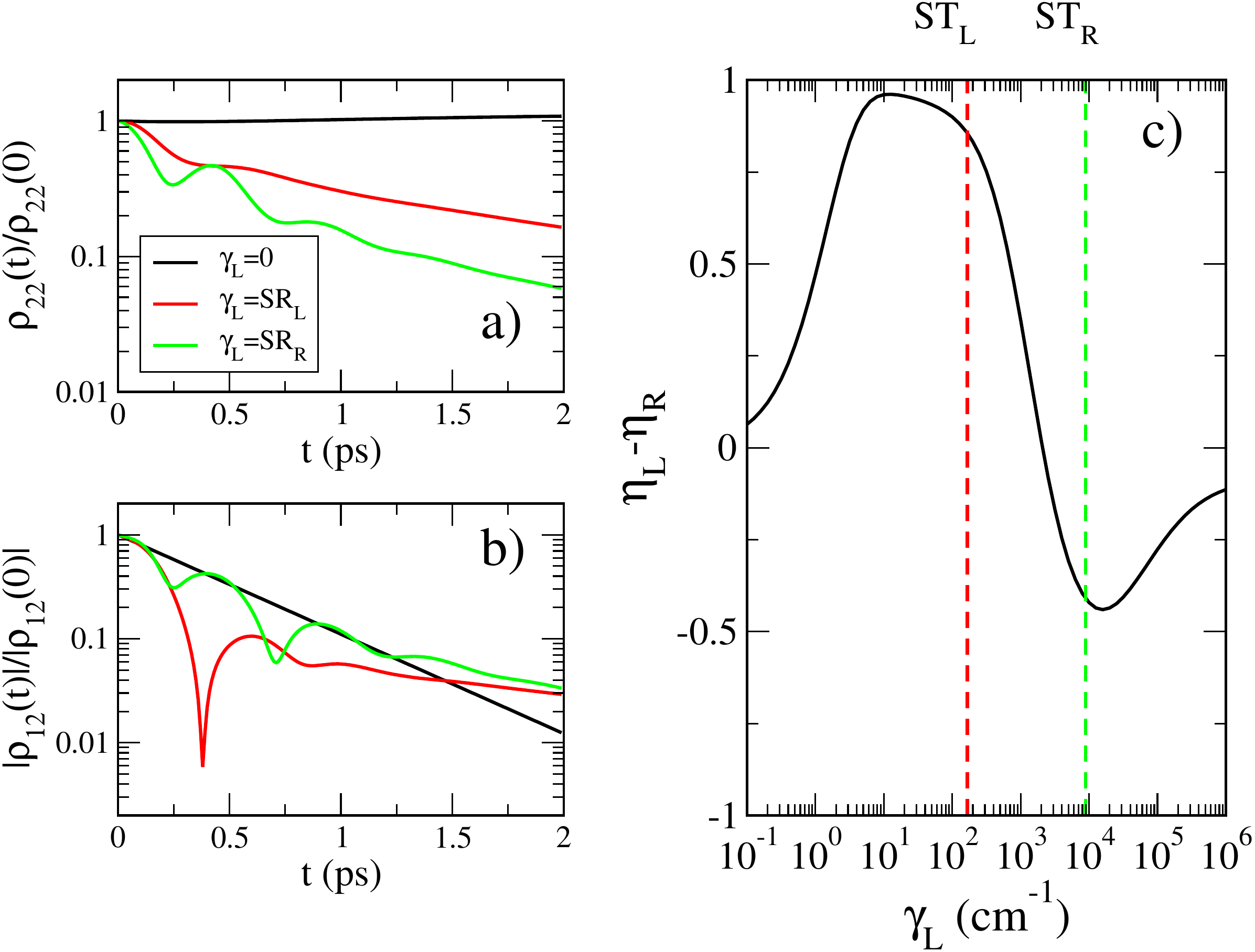}
\caption{(Color online) Realistic model of a RC. Left panels:  a) Decay of populations in time,
b) Decay of coherences in time for fixed ratio, $q=\gamma_{_L}/\gamma_{_R} = 100$,
and different values of $\gamma_{_L}$, as indicated in the legend.
Right panel c):  Unbalanced average left-right ET efficiency, $\eta_{_L} - \eta_{_R} $,
as a function of the effective coupling strength, $\gamma_{_L}$, for $300$ K.
A reorganization energy, $E_R= 20 \ cm^{-1}$,  and cut-off frequency, $\omega_c = 150 \ cm^{-1}$,
have been chosen  in order to have an approximate decay of populations and coherences
 on the time-scale of 1 ps. (See left panels a) and b.)
}
\label{ll4}
\end{figure}

\section{ Efficiency for a realistic model.}
Here  we apply our approach to a realistic model~ \cite{jenny} of the
photosystem II reaction center (PSII RC).
This system has  eight chromophores in the Left-Right subunits: two
chlorophylls belonging to the central “special pair”,
two accessory chlorophylls, two pheophytins
and two peripheral chlorophylls, not relevant since
weakly coupled.
To match  2D spectroscopy data, the energy levels in the active branch (left) are not
the same as for the inactive branch (right), and range between $15000$ and $15555$ $cm^{-1}$.
 Moreover, the coupling constants  among  chromophores are not equal for nearest neighbors,
and  all  chromophores couple to each other with  strengths varying  from $0.12$ to $162.2$ $cm^{-1}$.
The exact Hamiltonian matrix is not presented here. It appears in Ref.~\cite{jenny}.
Due to many differences from the simple model discussed above, one can wonder whether the previously discussed switching effect persists in this realistic model characterized by non-degenerate energy levels,
long-range interactions, and absence of exact left-right symmetry \cite{jenny}.

We proceed as before, by attaching the sinks to the pheophytins,   designing the effective non-Hermitian Hamiltonian, and calculating the conditions for the left and the right
STs (computed from the complex eigenvalues of the non-Hermitian Hamiltonian of this model,
 indicated as dashed vertical lines in Fig.~\ref{ll4}c). As one can see,
the maximal left/right ET efficiency roughly peaks near the $ST_{L,R}$.

In order  to show that the switch can work at room temperature,
we embedded the system in a thermal bath as described above.
We  note that the bath considered in Eq.~(\ref{eq:master-lindblad}),
without sinks,
produces typically  uncoupled equations for populations
(diagonal matrix elements)
and coherences (off-diagonal ones) in the energy basis.
In particular,
populations relax in time to Gibbs distribution  without oscillations,
at variance with the experiment~\cite{exp1,exp2,exp3,exp4,exp5,exp6,exp7,exp8,exp9}.
Interestingly, the coupling with the sinks
produces an effective coupling between populations and coherences.
The effect of the dynamics generated by the sinks is shown
in Figs.~\ref{ll4}a) and b), in which  oscillations are clearly
observable for $\gamma_{_L} \ne 0$, showing that coherences and populations
are now coupled.

We therefore use the same thermal bath as in Eq.~(\ref{eq:master-lindblad}), setting the parameters
for reorganization energy, $E_R$, and cut-off frequency, $\omega_c$, in order
to have  a decay of both populations and coherences of the order of   $1$ ps, in agreement
with experiments~\cite{exp8}.
Our results (shown in Fig.~\ref{ll4} c), demonstrate that  the previously observed switching survives in the presence of a thermal bath at room temperature (even though it  reduces the unbalanced  ET efficiency to the right branch ( $ \approx  0.5$ compared with the left  $ \approx 1$).

\section{ Conclusions.}
The analysis of our model for electron transfer, consisting of
two branches  attached to two asymmetric sinks,
revealed two different transport regimes: a classical one, in which the electron
transport occurs through the strongest coupled sink, and a quantum
one, in which the electron transport occurs through the weakest coupled sink.
Varying the coupling strengths in an appropriate way, see Fig.~\ref{phases}a,
one can switch from one regime to the other,
thus inducing a switching of electron transfer from one branch to the other.
This switching is a pure quantum effect,
based on
the existence of two consecutive superradiance transitions as the couplings vary.
The quantum nature of the switching is confirmed by the analysis of the
coupling to a thermal bath:
only quantum transport is weakened by the thermal bath, not the classical one.
This opens the possibility of using this switching effect to measure
the amount of quantum coherence in molecular networks.
Analyzing  a realistic model of the photosystem II reaction center at room temperature,
we have shown that this
 switching mechanism, being  robust to disorder and dephasing, could 
be observed in natural biological complexes.


\section{ Acknowledgments.}
This work has been supported by Regione Lombardia and CILEA Consortium through a LISA Initiative
(Laboratory for Interdisciplinary Advanced Simulation) 2011 grant [link:http://lisa.cilea.it].
 Support by the grant D.2.2 2011
(Calcolo ad alte prestazioni)  from Universit\'a  Cattolica
 is also acknowledged. The work by GPB and RTS was carried out under the auspices of the National Nuclear Security Administration of the U.S. Department of Energy at Los Alamos National Laboratory under
 Contract No. DE-AC52-06NA25396.  We also thank Gary Doolen for useful comments.


\newpage

\section{Appendix A:   Participation Ratio}

\begin{figure}[t!]
\includegraphics[scale=0.6]{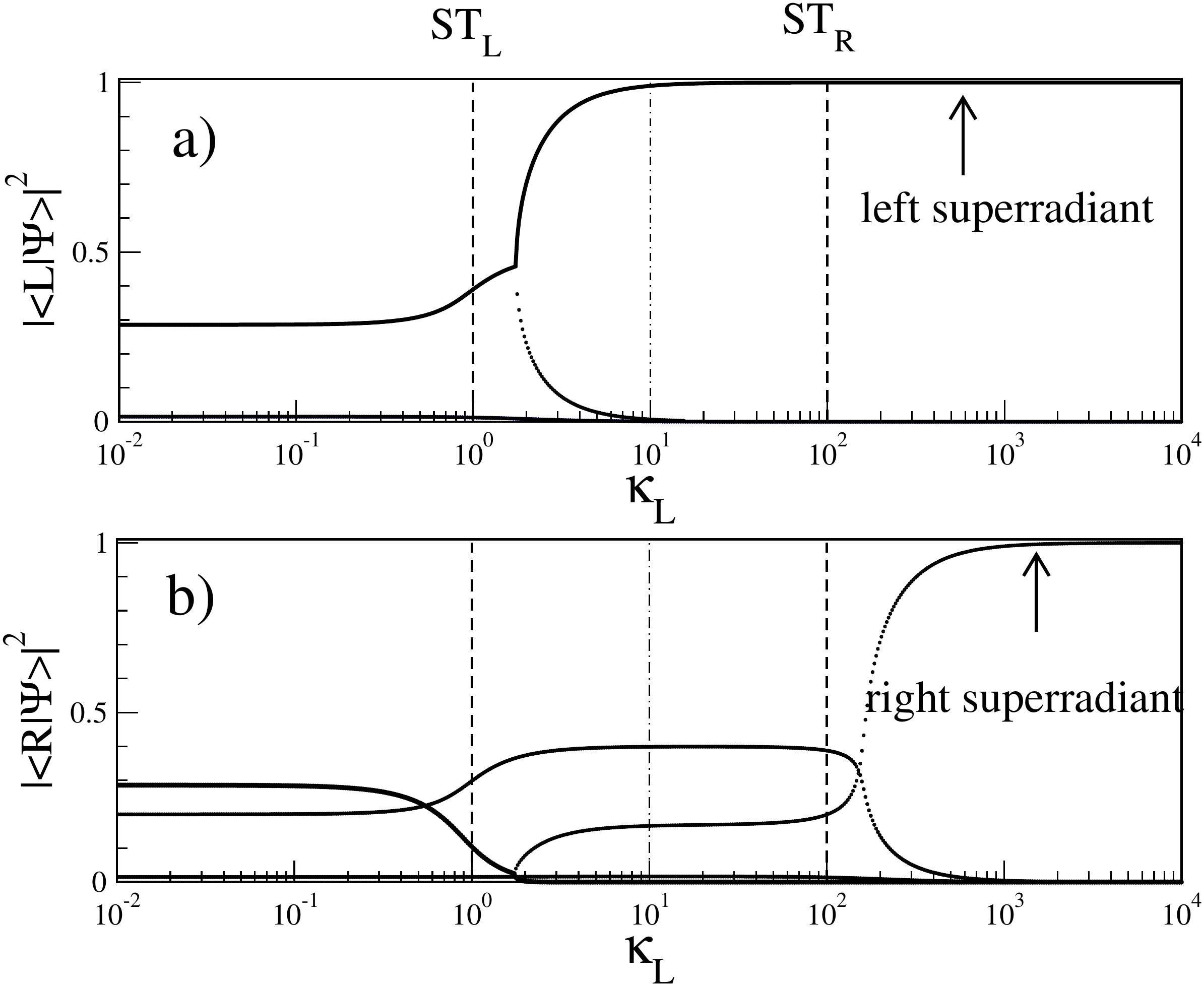}
\caption{a) Probability to be at the left end   $|\langle L| \Psi\rangle |^2$,
for all energy eigenstates $| \Psi \rangle$. b)
Probability to be at the right end   $|\langle R| \Psi\rangle |^2$,
for all energy eigenstates.
Here $q=100$.
}
\label{sup1}
\end{figure}
In Fig. \ref{sup1}a, we show the probabilities to occupy the end sites ($|L\rangle, |R\rangle$)
 for all energy eigenstates.
As one can see, on approaching  the left transition, $ST_L$, the superradiant state, $|SR\rangle$, increases  its probability to occupy
the left end, and this probability becomes one  at
the switching point.
Of course,  at the same point the probability to be at the left end
of the other states becomes zero, so that ET to the left sink is completely inhibited,
and ET switches to the right sink.

One can also observe that after the $ST_R$, the same effect occurs at very
large coupling thus inhibiting also the ET to the right end. This
is at variance with the classical behavior, where for very large coupling
one gets, $\eta_L = \eta_R = 1/2$.

\begin{figure}[t!]
\includegraphics[scale=0.6]{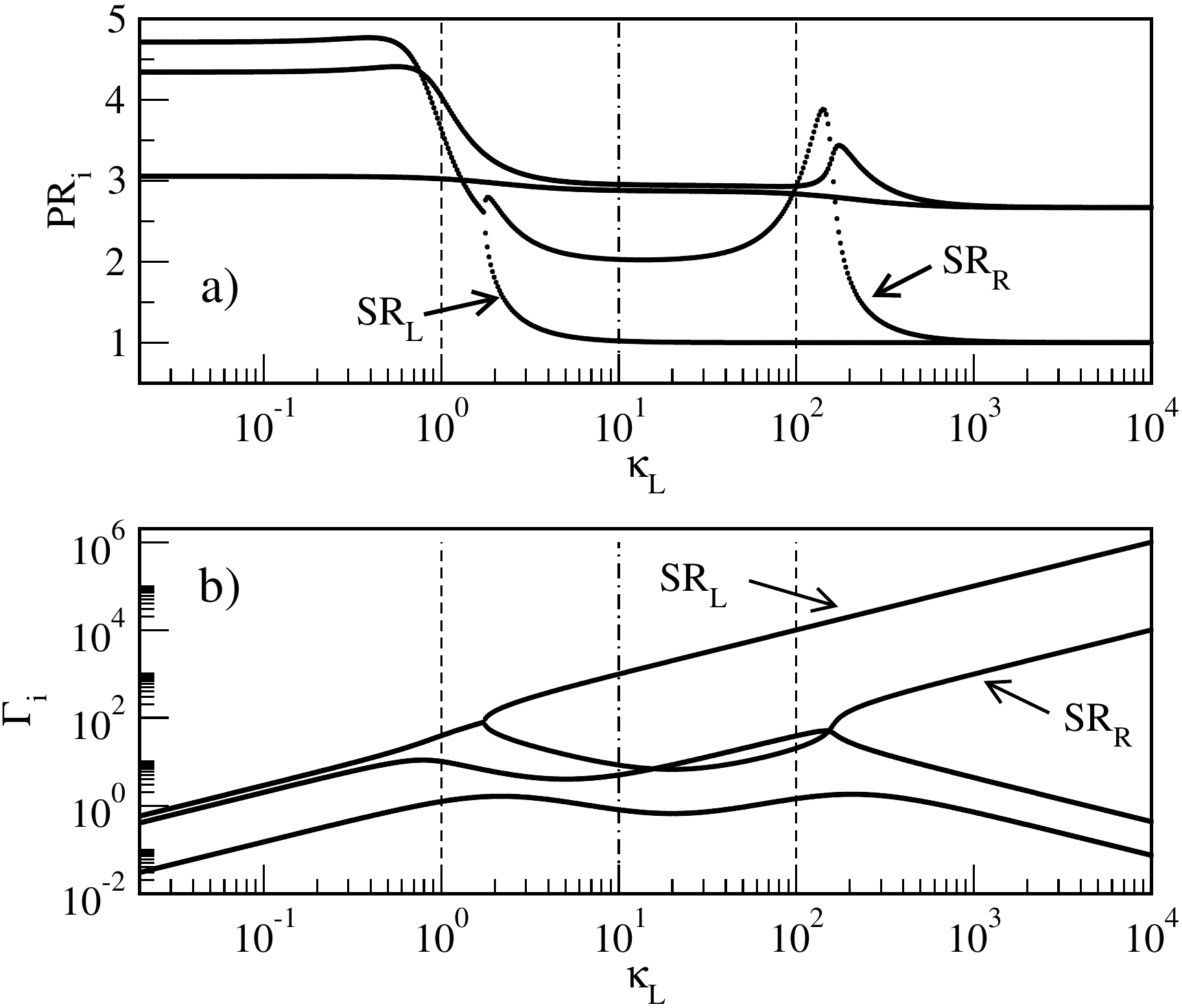}
\caption{a) Participation ratio versus $\kappa_L$  for all states.
b): Decay width versus  $\kappa_L$  for all states. Here $q=100$.
}
\label{sup2}
\end{figure}

In order to study the localization  of the eigenstates, $| \Psi \rangle$,
 we consider their
 participation ratio  in the site basis
$| n \rangle$,
$$
PR = \frac{1}{\sum _n | \langle n | \Psi \rangle|^4}.
$$
In Fig.~ \ref{sup2}a  we show the PR of all energy eigenstates.
As one can see, for   $ \kappa_L < 1$ most of the  states
are approximately
delocalized over the whole
system (the maximal $PR$  corresponds to the total number of sites, $6$).
For $1 <  \kappa_L < q$ both superradiant and subradiant states tend
to localize. However, while the $PR$ of the left superradiant state $SR_L$ becomes $ \approx 1$ at
the switching point $\kappa_L \sim \sqrt{ q} $  (complete
localization), the $PR$ of the other states does not decrease below $ \approx 3$, which
means that they are approximately extended over the system, thus allowing
the ET.

Even if the left superradiant state becomes extremely localized immediately after the $ST_L$,
its width is very large. (See Fig.~ \ref{sup2}b.) This competition between
localization and decay width determines the switching point of ET.

\section{Appendix B: The critical switching point}

Let us first define the left/right partial width as follow,
\beq
\label{pw}
\Gamma_{L,R} =     \gamma_{L,R} \ \sum_{k} \langle k | \  W_{L,R} \  | k \rangle,
\eeq
where the sum is taken over the subradiant states, $|k\rangle$.
Let us also consider the range of parameters between the two STs,
namely $1 < \kappa_L < q $, and the effective Hamiltonian,
\beq
\label{ssww1}
\mathcal{H} = -i\gamma_{_L}   W_L +  \left(   H_0  -i\gamma_{_R}   W_R  \right),
\eeq
where we consider $-i\gamma_{_L}   W_L $ as the  unperturbed Hamiltonian, and
$ H_0  -i\gamma_{_R}   W_R$, as  the perturbation.

Let us consider, for definiteness, the site $R$ as the first site $j=1$ and
the left $L$ as the last $j=N$.
Eigenvalues for $W_L$ are $1$ and $0$, the latter $N-1$ times degenerate;
the eigenvector correspondent to $1$ is $|L \rangle$, since,
$$
W_L | L\rangle = |L\rangle ,
$$
while, as a degenerate basis we can choose the site basis $|j\rangle$, for $j=1,...,N-1$,
since $W_L |j\rangle = 0 $.

Eigenvalues and eigenvectors of the $N-1$ degenerate system can be obtained
at zero order by solving the eigenvalues problem,
\beq
\langle k | H_0 -i\gamma_{_R} W_R | j \rangle = \epsilon_k \delta_{jk}.
\eeq
Since $H_0$ is of the order of $\Omega$, and $\gamma_{_R}/\Omega \ll 1 $, we can
use first order perturbation theory in $\gamma_{_R}/\Omega$  to get the eigenvalues,
\beq
\epsilon_q^{I} = \epsilon_q^{0} - i\gamma_{_R} \langle q | W_R |  q \rangle,
\eeq
and the eigenvectors,
\beq
|  q \rangle^{I} = |  q \rangle - \frac{i\gamma_{_R}}{2\Omega N} \sum_{q'\ne q}
\frac{\sin(q\pi/N) \sin(q'\pi/N)}{\sin[(q+q')\pi/N]\sin[(q-q')\pi/N]} |  q' \rangle
\equiv |  q \rangle - \frac{i\gamma_{_R}}{2\Omega N} \sum_{q'\ne q}
c_{q, q'}  |  q' \rangle,
\eeq
where we have chosen  the eigenbasis by the restriction of $H_0$ to the $N-1$
dimensional space,
$$
\langle k |  q \rangle = \sqrt{\frac{2}{N}} \sin \left( \frac{\pi k q }{N} \right),
$$
with eigenvalues $\epsilon_q^{0} = -2 \Omega \cos (\pi q /N) $.
The same eigenvectors can be considered (with the same name)
 in the $N$-th dimensional space, by simply
adding a $0$ in the $N$-th component, so to be orthogonal to $|L\rangle$.

To have the eigenstates at first order of the perturbation theory in $\Omega/\gamma_{_L}$ and
$\gamma_{_R}/\gamma_{_L}$,
one should take into account the interaction between $|  q \rangle$
and $|L\rangle$, mediated by the perturbation $H_0 -i\gamma_{_R} W_R$, so that,
\beq
\label{fopt}
|  q \rangle^{I} = |  q \rangle - \frac{i\gamma_{_R}}{2\Omega N} \sum_{q'\ne q}
c_{q, q'}  |  q' \rangle + \frac{ \langle q | H_0 -i\gamma_{_R} W_R |L\rangle }
{-i\gamma_{_L} - \epsilon_q^{0}}.
\eeq
From this one gets,
\beq
\begin{array}{lll}
\langle R | q \rangle^{I} &= \langle R | q \rangle + O(\gamma_{_R}/\Omega),\\
& \\
\displaystyle \langle L | q \rangle^{I} &=   i \sqrt{ \frac{2}{N}} \frac{\Omega}{\gamma_{_L}}
\sin \left(\frac{\pi q}{N}\right) + O( \Omega/\gamma_{_L} ).
\label{qfl5}
\end{array}
\eeq
From Eq.~(\ref{qfl5}) the partial widths easily follows,
\beq
\begin{array}{lll}
\displaystyle \Gamma_R & = \frac{2\gamma_{_R}}{N}  \sum_q     \sin^2(\pi q/N), \\
& \\
\displaystyle \Gamma_L & = \frac{2\Omega^2}{N\gamma_{_L}}  \sum_q     \sin^2(\pi q/N).
\label{qfl6}
\end{array}
\eeq
Equating Eq.~(\ref{qfl5}) and Eq.~(\ref{qfl6}), one gets,
$$
\gamma_{_L} \gamma_{_R} \simeq \Omega^2.
$$

\section{Appendix C: Lindblad master Equation}

The dynamics of the system to second order in the
system-bath coupling can be described by the Lindblad
master equation in the Born-Markov and secular approximations
as
\begin{equation}
\label{eqlindblad}
\frac{d\rho}{dt} = -\frac{i}{\hbar} \left( \mathcal{H}\rho -\rho\mathcal{H}^{\dagger} \right)
 + L_p(\rho),
\end{equation}
where the superoperator, $L_p$, acts on $\rho$ as follow,
\beq
\begin{array}{lll}
L_p(\rho) &= \sum_{\omega,m}  \gamma(\omega) [ A_m(\omega) \rho A_m^\dagger (\omega)
-\frac{1}{2} A_m^\dagger (\omega) A_m(\omega) \rho \\
& \\
&-\frac{1}{2} \rho  A_m^\dagger (\omega) A_m(\omega)].
 \end{array}
\label{sdey}
\eeq
The Lindblad generators, $A_m(\omega)$, are given by,
\begin{equation}
\label{supa}
 A_m(\omega) = \sum_{E-E'= \hbar \omega} c_m^* (E) c_m(E') |E \rangle \langle E' |,
\end{equation}
where the summation is over all transitions with frequency,  $\omega = (E-E')/\hbar$,
and $|  E \rangle$ is the eigenstate of the Hamiltonian, $H_0$, of
the closed system  with eigenvalue, $E$. The coefficients, $c_m(E)$, are the expansion coefficients of the energy eigenstate in the sites
basis, $| m \rangle$,
$$
|E\rangle = \sum_m c_m (E) |m\rangle.
$$

\noindent
The rates, $\gamma (\omega)$, are given by,
$$
\gamma(\omega) = 2\pi [ J(\omega)(1 + n_T(\omega) ) + J(-\omega) n_T(-\omega)],
$$
where $n_T(\omega)$ is the bosonic distribution function at the temperature $T$,
$$
n_T(\omega ) = \frac{1}{ e^{\hbar \omega/k_B T } -1},
$$
and $J(\omega)$ is the Ohmic spectral density, which we choose of the form,
\beq
 J(\omega)  = \left\{
\begin{array}{lll}
     & 0  \quad &{\rm for  } \quad \omega < 0 \\
     & \displaystyle
 \frac{E_R \omega}{\hbar \omega_c} e^{-\omega/\omega_c}  \quad   &{\rm for} \quad  \omega > 0.
 \end{array}
\right.
\label{sde}
\eeq
Eq.~(\ref{sde}) implicitly  defines the reorganization energy, $E_R$, and the cut-off
frequency, $\omega_c$.

Note that this form of master equation does not assume
weak coupling to the sinks, only to the phonon bath.
We also assumed that a strong coupling
to the sinks does not influence the phonon coupling.
Further work is in progress to check the validity
of this assumption.

\end{document}